\documentclass[twocolumn]{jpsj3}

\usepackage{color}
\usepackage{times}
\usepackage{graphicx}

\title{Chaos in Jahn-Teller Rattling}

\author{Takashi Hotta and Akira Shudo}

\inst{Department of Physics, Tokyo Metropolitan University,
Hachioji, Tokyo 192-0397, Japan}

\recdate{\today}

\abst{
We unveil chaotic behavior hidden in the energy spectrum of
a Jahn-Teller ion vibrating in a cubic anharmonic potential
as a typical model for rattling in cage-structure materials.
When we evaluate the nearest-neighbor level-spacing distribution
$P(s)$ of eigenenergies of the present oscillator system,
we observe the transition of $P(s)$ from the Poisson to the Wigner distribution
with the increase of cubic anharmonicity,
showing the occurrence of chaos in the anharmonic Jahn-Teller vibration.
The energy scale of the chaotic region is specified from
the analysis of $P(s)$ and we discuss a possible way to observe chaotic behavior
in the experiment of specific heat.
It is an intriguing possibility that chaos in nonlinear physics could be detected
by a standard experiment in condensed matter physics.
}

\kword{
Chaos, Rattling, Jahn-Teller vibration, Cubic anharmonicity
}

\begin{document}
\maketitle


Recently, a peculiar magnetically robust heavy-electron phenomenon
observed in Sm-based filled skutterudite compound \cite{Sanada}
has triggered active investigations on cage-structure compounds,
in which a guest ion contained in a cage composed of relatively
light atoms oscillates with large amplitude in an anharmonic potential.
Such a local vibration with large amplitude is called {\it rattling} and
exotic magnetism and superconductivity induced by rattling have attracted
much attention in the research field of condensed matter physics.
As easily understood from the above explanation, rattling is considered
to be one of typical nonlinear phenomena, but unfortunately,
such a viewpoint has not been recognized at all in the research field
of nonlinear physics.

However, we emphasize an important point of contact between
rattling phenomena and nonlinear physics
through a concept of {\it chaos}.
For nonlinear physicists, it will be quite natural to expect
the appearance of chaos in the two-dimensional oscillator
in a potential with plural numbers of minima.
In fact, such a situation is expected to occur in cage-structure compounds,
if we consider a Jahn-Teller ion vibrating in a cubic anharmonic potential,
\cite{Hotta}
in the course of the research of the Kondo effect with phonon origin.
\cite{Kondo1,Kondo2,Vladar1,Vladar2,Vladar3,Yu-Anderson,Matsuura1,Matsuura2,
Yotsuhashi,Hattori1,Hattori2,Mitsumoto1,Mitsumoto2,
Hotta1,Hotta2,Hotta3,Hotta4,Hotta5,Hotta6,
Yashiki1,Yashiki2,Yashiki3,Hattori3,Hotta7,
Fuse1,Fuse2,Fuse3,Fuse4,Fuse5,Fuse6}
It is worth to point out that such a cubic anharmonic term of Jahn-Teller
vibration just indicates the H\'enon-Heiles potential which
has been discussed for the appearance of chaos
in the early stage.\cite{HH}

Actually, apart from the rattling problem,
the appearance of chaos in the vibronic state,
i.e., the complicated electron-vibration coupled state,
has been already pointed out by several groups.
\cite{Cederbaum,JTC0,JTC1a,JTC1b,JTC2,JTC3,JTC4}
However,  we hit upon an idea that
{\it chaos originates from anharmonic Jahn-Teller vibration},
not from the vibronic state
composed of electron and anharmonic Jahn-Teller vibration.
It is important to confirm that the origin of chaos exists
in the anharmonic Jahn-Teller oscillator, since it will provide
us a realistic model in condensed-matter physics
for the research of chaos in nonlinear physics.

In this Letter, we clarify the chaotic property in
anharmonic Jahn-Teller vibration.
In order to confirm the appearance of chaos,
we evaluate the nearest-neighbor level-spacing distribution $P(s)$,
indicating that $P(s)$ changes from the Poisson to the Wigner distribution
with the increase of cubic anharmonicity.
We also discuss the energy region in which chaotic behavior occurs.
In order to consider a possible way to detect the chaotic behavior,
we propose the measurement of specific heat in cage-structure materials.
It is pointed out that the peak structure in the temperature dependence
of specific heat could be a signal of chaotic behavior.

Let us consider a Jahn-Teller oscillator in an anharmonic potential.
The Hamiltonian is given by \cite{unit}
\begin{equation}
  H=(P_1^2+P_2^2)/(2M)+V(Q_1,Q_2),
\end{equation}
where $M$ is the reduced mass of Jahn-Teller oscillator,
$Q_1$ and $Q_2$ denote normal coordinates of 
$(3z^2-r^2)$- and $(x^2-y^2)$-type Jahn-Teller oscillation, respectively,
$P_1$ and $P_2$ indicate corresponding canonical momenta, and
$V(Q_1,Q_2)$ is the potential for the Jahn-Teller oscillator.
The potential is given by
$V(Q_1, Q_2)$=$A(Q_1^2+Q_2^2)$+$B(Q_1^3-3 Q_1 Q_2^2)$+$C(Q_1^2+Q_2^2)^2$,
where $A$ is the quadratic term of the potential,
while $B$ and $C$ denote the coefficients for third- and fourth-order
anharmonic terms, respectively.
As mentioned above, the third-order term is just
the H\'enon-Heiles potential.\cite{HH}
Note also that we include only the anharmonicity which maintains
the cubic symmetry.
Among the coefficients, $A$ and $C$ are taken as positive,
while $B$ is set as negative in this research.

In order to understand the properties of the potential,
it is convenient to introduce the non-dimensional distortion as
$q_1$=$\sqrt{2M\omega}Q_1$ and $q_2$=$\sqrt{2M\omega} Q_2$,
where $\omega$ is the phonon energy given by $\omega$=$\sqrt{2A/M}$.
By introducing $q$ and $\theta$ through the relations of
$q_1$=$q \cos \theta$ and $q_2$=$q \sin \theta$,
we obtain $V$ as
\begin{equation}
  \label{eq:pot}
  V(q, \theta)=\omega (q^2/4+\beta q^3 \cos 3\theta/3
                    +\gamma q^4/8),
\end{equation}
where non-dimensional anharmonicity parameters are defined by
$\beta$=$3B/[(2M)^{3/2}\omega^{5/2}]$ and 
$\gamma$=$2C/(M^2\omega^3)$.
Note that the energy scale of the potential is given by $\omega$.
Thus, in the following, we set the energy unit as $\omega$=$1$.

In this potential, for $|\beta|$$\le$$\sqrt{\gamma}$,
there is a single minimum at $q$=$0$,
while for $|\beta|$$>$$\sqrt{\gamma}$,
there appear three minima for $q$$\ne$$0$
in addition to the shallow minimum at $q$=$0$.
In Fig.~1(a), we plot $V(q,\theta)$ vs. $q$
for several values of $\theta$
for the case of $\beta$=$-2$ and $\gamma$=$1$.
Along the direction of $\theta$=$0$, we find a deep minimum,
while we find a saddle point along the direction of $\theta$=$\pi/3$.
The potential structure is gradually changed with the increase of
$\theta$.
In Fig.~1(b), we show  the contour plot of the potential $V$
for $\beta$=$-2$ and $\gamma$=$1$.
Here we find three minima along the directions of
$\theta$=$0$, $2\pi/3$, and $4\pi/3$,
corresponding to $(3z^2-r^2)$-, $(3x^2-r^2)$-, and $(3y^2-r^2)$-type
Jahn-Teller distortions, respectively.
Note that there still remains trigonal symmetry, as easily understood from
the term of $\cos 3\theta$ in eq.~(\ref{eq:pot}),
since we consider the cubic anharmonicity.

\begin{figure}[t]
\centering
\includegraphics[width=8.0truecm]{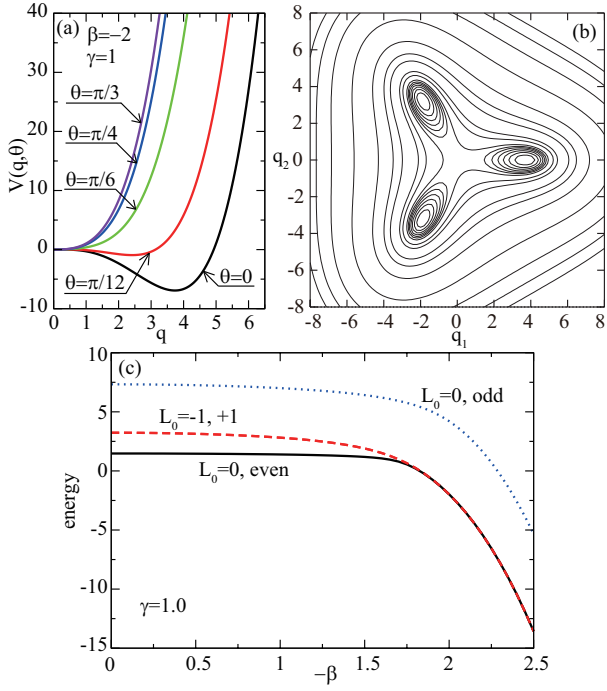}
\caption{(Color online)
(a) Potential $V$ vs. $q$ along the directions of
$\theta$=$0$, $\pi/12$, $\pi/6$, $\pi/4$, and $\pi/3$
for $\beta$=$-2$ and $\gamma$=$1$.
(b) Contour plot of $V$ on the $q_1$-$q_2$ plane
for $\beta$=$-2$ and $\gamma$=$1$.
Here we draw the contour curves for
$V$=$-6$, $-5$, $-4$, $-3$, $-2$, $-1$, $0.5$, $5$, $10$,
$30$, $50$, $100$, $250$, $500$, and $750$.
(c) Eigenenergies vs. $-\beta$ for the states of $L_0$=$0$ and $\pm 1$.
Note that the states with $L_0$=$0$ is further classified into two types,
specified by ``even'' and ``odd''.
}
\end{figure}

For the later discussion, here we define the potential depth $V_0$ as
$V_0$=$V(q_-,\theta=0)$$-$$V(q_+,\theta=0)$,
where $q_{\pm}$ denotes the position of extrema,
given by $q_{\pm}$=$(-\beta \pm \sqrt{\beta^2-\gamma})/\gamma$.
Then, we obtain $V_0$ as
\begin{equation}
 \label{eq:v0}
  V_0=2|\beta|(\beta^2-\gamma)^{3/2}/(3\gamma^3).
\end{equation}
Note that $V_0$ is defined for $|\beta|$$\ge$$\sqrt{\gamma}$.

In order to discuss the local phonon state,
it is necessary to perform the quantization procedure
through the relations of $q_1$=$a_1$+$a_1^{\dag}$
and $q_2$=$a_2$+$a_2^{\dag}$,
where $a_1$ and $a_2$ are annihilation operators of
phonons for Jahn-Teller oscillations.
In order to unveil the conserved quantities in the Hamiltonian $H$,
it is useful to introduce
the transformation of phonon operators as
$a_{\pm}$=$(a_1 \pm {\rm i}a_2)/\sqrt{2}$,\cite{Takada}
where the sign in this equation intuitively indicates
the rotational direction in the potential.
With the use of these operators, the Hamiltonian is rewritten as
\begin{equation}
  \begin{split}
   H & = a_{+}^{\dag}a_{+}+a_{-}^{\dag}a_{-}+1 \\
     & +(\beta/3) [(a_{+} + a_{-}^{\dag})^3
       +(a_{-} + a_{+}^{\dag})^3] \\
     & +(\gamma/2) (a_{+}^{\dag}a_{+}+a_{-}^{\dag}a_{-}+1
       +a_{+}^{\dag}  a_{-}^{\dag}+a_{+}a_{-})^2.
  \end{split}
\end{equation}
Note again that the energy unit is set as $\omega$=$1$.

In order to diagonalize the Hamiltonian,
we prepare the phonon basis $|L; n\rangle$, given by
\begin{equation}
|L; n \rangle = \left \{
\begin{array}{ll}
|L+n, n \rangle & L \ge 0 \\
|n, n+|L| \rangle & L <0,
\end{array}
\right.
\end{equation}
where the phonon basis $|n_+,n_- \rangle$ is given by
$|n_+,n_- \rangle$=
$(1/{\sqrt{n_+! n_-!}})
(a_+^{\dag})^{n_+}(a_-^{\dag})^{n_-}|0\rangle$
with the vacuum $|0\rangle$.
In actual numerical calculations
to solve the eigenvalue problem,
the phonon basis $|L; n \rangle$ is truncated
at a finite number $N_{\rm ph}$
and a maximum angular momentum $L_{\rm max}$.
In order to check the convergence,
we have performed the numerical calculations for $N_{\rm ph}$
and $L_{\rm max}$ up to 250 and 125, respectively.

For $\beta$=$0$, we find that the quantum phonon state is
labelled by the angular momentum $L$, given by
$L$=$a_{+}^{\dag}a_{+}$$-$$a_{-}^{\dag}a_{-}$.
Note that $L$ commutes with $H$ for $\beta$=$0$.
Thus, when we diagonalize the Hamiltonian for $\beta$=$0$,
we prepare the phonon basis for a fixed value of $L$,
since the states with different $L$ are not mixed.
When the potential has continuous rotational symmetry for $\beta$=$0$,
the angular momentum should be the conserved quantity in
the quantum mechanics.

When we include the effect of $\beta$, namely, cubic anharmonicity,
the situation is changed.
As easily understood from eq.~(\ref{eq:pot}),
there occurs the trigonal term in the potential.
In such a case, $L$ is no longer the good quantum number,
but there still exists conserved quantity concerning $L$.
In order to clarify such a point, we express $L$ as
\begin{equation}
  L=3\ell+L_0,
\end{equation}
where $L_0$ takes the values of $0$ and $\pm 1$.
It is found that $L_0$ is the good quantum number.

The states with $L_0$=$\pm 1$ is expressed by
\begin{equation}
|\Phi^{(\pm 1)}_{k} \rangle
=\sum_{\ell,n} \varphi_{\ell,n}^{(k,\pm 1)} |3\ell \pm 1; n\rangle,
\end{equation}
where $|\Phi^{(L_0)}_{k} \rangle$ denotes the $k$-th eigenstate
characterized by quantum number $L_0$
and $\varphi$ is the coefficient of the eigenstate.
The corresponding eigenenergy is expressed by
$E^{(L_0)}_{k}$.

Note that for the case of $L_0$=$0$,
there exists extra conserved quantity of parity,
concerning the change of $\ell \rightarrow -\ell$.
It is simply understood from the fact that
the bonding and anti-bonding states of
$|3\ell;n\rangle$ and $|-3\ell;n\rangle$
are not mixed with each other.
Then, the parity for the bonding (even) or anti-bonding (odd) state
is another good quantum number.
Note that the bonding state of $|3\ell;n\rangle+|-3\ell;n\rangle$
is mixed with $|0;n\rangle$, but the anti-bonding state
$|3\ell;n\rangle-|-3\ell;n\rangle$ is not.

The state for $L_0$=$0$ with even parity is given by
\begin{equation}
\begin{split}
|\Phi^{(0{\rm e})}_{k} \rangle
 &= \sum_n \varphi_{0,n}^{(k,0{\rm e})} |0; n\rangle \\
 &+\sum_{\ell>0,n} \varphi_{\ell,n}^{(k,0{\rm e})}
(|3\ell; n\rangle +|-3\ell; n\rangle)/\sqrt{2},
\end{split}
\end{equation}
while the state for $L_0$=$0$ with odd parity is given by
\begin{equation}
|\Phi^{(0{\rm o})}_{k} \rangle
=\sum_{\ell>0,n} \varphi_{\ell,n}^{(k,0{\rm o})}
(|3\ell; n\rangle -|-3\ell; n\rangle)/\sqrt{2}.
\end{equation}
In short, we classify the eigenstates for the case of $\beta$$\ne$$0$
into four groups with the labels of
``$+1$'', ``$-1$'', ``$0$e'', and  ``$0$o''.

In Fig.~1(c), we plot $E^{(0{\rm e})}_{0}$, $E^{(\pm 1)}_{0}$, 
and $E^{(0{\rm o})}_{0}$, which denote the lowest eigenenergies
of the groups of ``$0$e'', ``$\pm 1$'', and ``$0$o'', respectively.
We observe that the ground state is always given by
$|\Phi^{(0{\rm e})}_{0} \rangle$ and the first excited state
is doubly degenerate, given by $|\Phi^{(\pm 1)}_{0} \rangle$.
Note that these three states seem to be almost degenerate
in the region of $\beta$$<$$-1.7$.
The state of $|\Phi^{(0{\rm o})}_{0} \rangle$ appears
in the relatively high-energy region.

The lowest-energy state of $L_0$=$0$ with even parity
includes the significant contribution of $L$=$0$
and it corresponds to the zero-point oscillation.
On the other hand, the state with $L_0$=$\pm 1$ has excitation
with finite angular momentum.
Around at $\beta$$\approx$$-1.7$, the zero-point energy
is found to be less than the potential depth $V_0$,
suggesting that the oscillation states begin to be
localized in the potential minima.
In such a situation, the potential minima become deep
and the quantum tunneling among potential minima
is suppressed.
Thus, the energy difference due to rotational motion
becomes very small and the excitation energy is extremely
reduced.

\begin{figure}[t]
\centering
\includegraphics[width=8.0truecm]{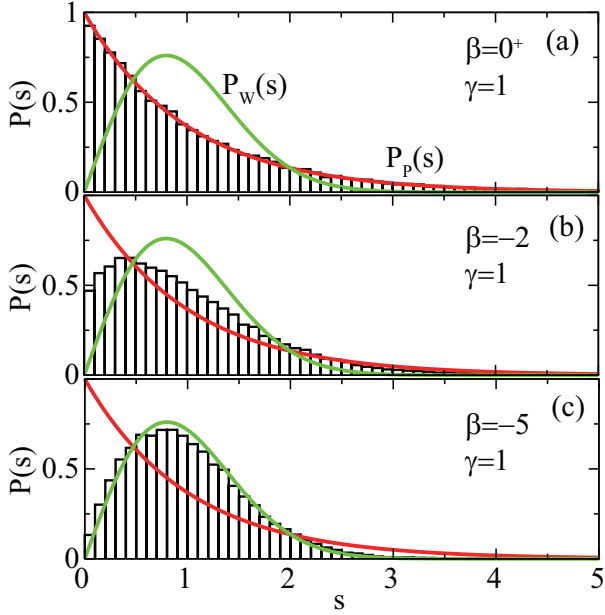}
\caption{(Color online)
Nearest-neighbor level-spacing distribution $P(s)$ for
(a) $\beta$=$0^+$, (b) $\beta$=$-2$, and (c) $\beta$=$-5$
for $\gamma$=$1$.
Note that  $P_{\rm P}(s)$ and $P_{\rm W}(s)$ denote
the Poisson and Wigner distributions,  respectively, given by
$P_{\rm P}(s)$=$e^{-s}$ and
$P_{\rm W}(s)$=$(\pi s/2)e^{-\pi s^2/4}$.
To draw $P(s)$, we use the numerical data for $L_0$=$1$,
but there is no significant difference in $P(s)$
even for $E_k^{(-1)}$ or $E_k^{(0{\rm e})}$.
}
\end{figure}

Now we explain a way to extract information on chaos from
the energy spectrum.\cite{Bohigas}
First we prepare the eigenenergies $\{ E_k \}$
for each quantum number.
Note that we cannot obtain correct distribution,
if the eigenstates with different symmetry are mixed.
Next we introduce the average counting function $\langle N(E) \rangle$, 
where $N(E)$ denotes the number of energy levels less than $E$, 
and perform the procedure of ``unfolding'' by
the mapping $x_k$=$\langle N(E_k) \rangle$
with the unfolded level $x_k$.
Then, we evaluate the distribution of nearest-neighbor level-spacing
$P(s)\Delta s$ by counting the number of spacings satisfying 
$s$$<$$x_k$$-$$x_{k-1}$$<$$s$$+$$\Delta s$
with an appropriate mesh $\Delta s$.

In Fig.~2, we show $P(s)$ obtained from $E_k^{(+1)}$
with $\Delta s$=$0.1$ for $\beta$=$0^{+}$, $-2$, and $-5$.
Here $0^{+}$ indicates the infinitesimal small positive number.
Note that there is no significant difference in the distribution $P(s)$,
if we use the numerical data of
$E_k^{(-1)}$ and $E_k^{(0{\rm e})}$.
For $\beta$=$0^+$, we observe the Poisson distribution
$P_{\rm P}(s)$=$e^{-s}$,
while for $\beta$=$-5$, we find the Wigner distribution
$P_{\rm W}(s)$=$(\pi s/2)e^{-\pi s^2/4}$.
For $\beta$=$-2$, the mixture of $P_{\rm P}(s)$ and
$P_{\rm W}(s)$ is observed.
It is well known that 
when the classical system exhibits chaos, 
eigenenergies of
the corresponding quantum system shows the Wigner distribution.
Thus, we conclude that the chaotic behavior appears
when we increase the cubic anharmonicity $\beta$.
We also emphasize that the chaotic behavior observed
in the vibronic state in the previous research should be
considered to originate from chaos in
the anharmonic Jahn-Teller vibration.

\begin{figure}[t]
\centering
\includegraphics[width=8.0truecm]{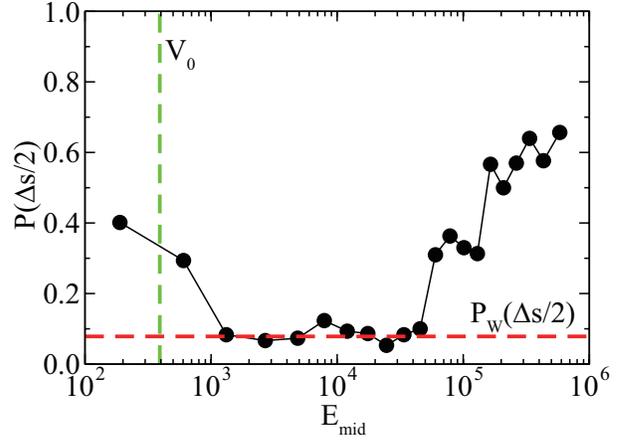}
\caption{(Color online)
$P(\Delta s/2)$ vs. $E_{\rm mid}$ for
$\beta$=$-5$ for $\gamma$=$1$.
We set $\Delta s$=$0.1$ in the evaluation of $P(s)$.
As for the meaning of $E_{\rm mid}$, see the maintext.
A vertical line denotes the potential depth $V_0$,
while a horizontal line indicates $P_{\rm W}(\Delta s/2)$.
}
\end{figure}

Here we have a naive question:
In which energy region chaos predominates ?
In order to reply to this question, we divide the sequence of eigenenergies
into several sectors and evaluate $P(s)$ of each sector.
We consider the eigenenergies with $L_0$=$+1$
for $\beta$=$-5$ and $\gamma$=$1$,
where $V_0$=$160\sqrt{6}$=$392$ from eq.~(\ref{eq:v0}).
We prepare the following sequence of integer:
$k_0$=$0$, $k_1$=$550$, and $k_j$=$3000$$\times$$(j-1)$
for $j \ge 2$.
Note that $k_1$ is determined so as to satisfy the relation of
$E^{(+1)}_{k_1}$=$V_0$.
We define the sector $j$ including the eigenenergies
from $k_{j-1}$ to $k_j$.
Then, for each sector $j$, we evaluate $P(s)$.

In Fig.~3, we plot $P(\Delta s/2)$ vs. $E_{\rm mid}$
for $\Delta s$=$0.1$, where $E_{\rm mid}$ indicates
the energy just at the center of the sector $j$.
The horizontal line denotes the value of $P_{\rm W}(\Delta s/2)$.
We find that for small $E_{\rm mid}$ comparable with $V_0$,
$P(\Delta s/2)$ is apparently larger than $P_{\rm W}(\Delta s/2)$,
suggesting that the chaotic nature is weak.
This is understood from the fact that the oscillation is harmonic
near the bottom of the potential well, since the potential is quadratic
near the potential minimum.
For large $E_{\rm mid}$, we also observe that 
$P(\Delta s/2)$ significantly deviates from $P_{\rm W}(\Delta s/2)$.
For the energy region much larger than $V_0$, the potential
is dominated by the fourth-order term with $q^4$ and
the system asymptotically approaches the integrable system,
suggesting that the chaotic nature should disappear.

In the energy region for moderately larger than $V_0$,
we find the relation of $P(\Delta s/2)$$\approx$$P_{\rm W}(\Delta s/2)$,
although some deviations occur due to the statistical property.
In one word, the result clearly defines the energy region
with chaotic behavior.
Namely, the chaotic nature comes from the eigenenergies
between $a_{\rm L}V_0$ and $a_{\rm H}V_0$,
where $a_{\rm L}$ is a number of the order of unity,
while $a_{\rm H}$ is in the order of hundred.
It seems natural that the energy of the chaotic region is related with
the potential depth $V_0$, but such energy region spreads over
a few hundred times larger than $V_0$.
The width of the chaotic region is much larger
than we have naively expected.

\begin{figure}[t]
\centering
\includegraphics[width=8.0truecm]{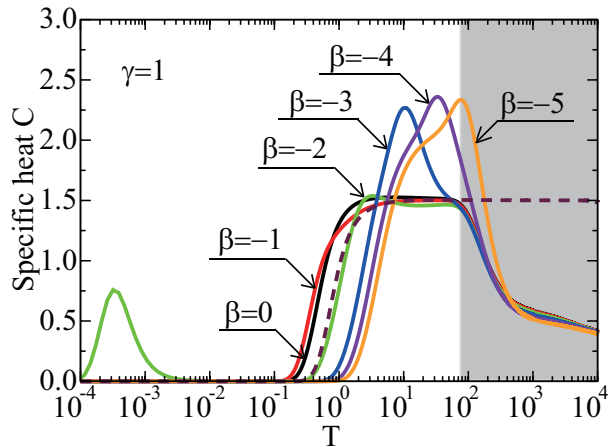}
\caption{(Color online)
Specific heat $C$ vs. $T$ for several values of $\beta$
with $\gamma$=$1$.
The broken curve indicates the result of the one-dimensional
anharmonic oscillator.
A shaded square indicates a region in which
numerical results for $C$ did not converge satisfactorily
in the present calculations.
}
\end{figure}

Now we discuss a possible way to detect the chaotic nature
in observables. 
For the purpose, we evaluate the specific heat $C$, given by
$C$=$(\langle H^2 \rangle-\langle H \rangle^2)/T^2$,
where $T$ is a temperature and 
$\langle H^m \rangle$=$\sum_k e^{-E_k/T} E_k^m/Z$
with the partition function $Z$=$\sum_k e^{-E_k/T}$.
Since the evaluation of $C$ is done by the numerical calculation
with the use of finite numbers of phonon bases,
we should note that $C$ may exhibit unphysical behavior
at high temperatures.

In Fig.~4, we show the specific heat $C$ vs. temperature $T$
for several values of $\beta$.
Note that the unit of $C$ is $k_{\rm B}$, which is set as unity.
\cite{unit}
For $\beta$=$0$ and $-1$, $C$ is increased rapidly around
at $T$$\sim$$1$ and it becomes almost a constant value,
corresponding to the Dulong-Petit law.
Note that the result agrees well with the broken curve of $2C_1$
in the high-temperature region, where $C_1$ denotes
the specific heat of one-dimensional anharmonic oscillator
in the potential of $V_1(q)$=$\omega(q^2/4$$+$$q^4/8)$
with non-dimensional length $q$.
At high enough temperatures,
all the results should approach
the broken curve, since the potential is dominated
by the fourth-order term in the high-energy region.
However, in the actual calculations with finite numbers of phonon bases,
it is inevitable that $C$ is deviated from the constant value
at some temperature,
as denoted by a shaded region in Fig.~4.

For $\beta$=$-2$,  we find a hump at $T$$\sim$$3.3$.
At low temperatures, we find a Schottky peak
determined by the first excitation energy
$\Delta E$=$E_0^{(\pm 1)}$$-$$E_0^{(0{\rm e})}$,
which is the difference between lowest two curves in Fig.~1(c).
Note that the Schottky peak cannot be observed for $\beta$=$0$ and $-1$,
since $\Delta E$ is larger than unity for both cases.
On the other hand, for $\beta$$<$$-2$, the Schottky peak exists,
but the peak position is smaller than $10^{-4}$.

With the increase of $|\beta|$, the hump found in $\beta$=$-2$ grows
and it eventually becomes the robust peak structure.
The temperature at the peak $T_{\rm p}$ is found to be given by
$T_{\rm p}$=$11$, $33$, and $77$
for $\beta$=$-3$, $-4$, and $-5$, respectively.
These values are well scaled by $V_0$ in eq.~(\ref{eq:v0}).
For large values of $|\beta|$ such as $\beta$=$-4$ and $-5$,
we have carefully checked that the value of $T_{\rm p}$ converges
even in the present numerical calculations, although it is difficult
to reproduce the Dulong-Petit law consistent with $2C_1$
in the high-temperature region.
Note, however, that we observe a shoulder in the position of $2C_1$
for $\beta$=$-3$.

From Figs.~3 and 4, the appearance of the peak structure
with large width in the specific heat over the broken curve
in the high-temperature region seems to characterize
the chaotic nature,
since we intuitively consider that an entropy is expected to be
enhanced due to 
uniform spreading of the eigenfunctions in the phase space
for the energy region with chaotic behavior.\cite{Berry,Voros}
Here we note that $C$ is related with the entropy $S$ as
$C$=$T (\partial S/\partial T)$,
suggesting that $C$ forms a peak when the entropy
is rapidly increased with the increase of $T$.
Thus, we deduce that
the robust peak structure in the specific heat becomes
a signal of the emergence of chaos.

In the experiments of cage-structure materials,
the specific heat has been usually measured.
In many cases, experimentalists have plotted
the value of $C/T^3$ as a function of $T$,
since the Debye specific heat is in proportion to $T^3$
at low temperatures.
In the plot of $C/T^3$ vs. $T$, we obtain the peak structure
corresponding to the characteristic frequency of
the Einstein phonon for rattling.
However, our proposal is to seek for the peak in $C$, not in $C/T^3$,
as the enhancement of $C$ due to the chaotic nature
of anharmonic Jahn-Teller vibration.
A candidate material is cage-structure compound
with off-center rattling such as clathrate.
Note that we consider the specific heat
in the temperature region higher than a room temperature.

In summary, we have clarified the chaotic nature of Jahn-Teller rattling.
From the evaluation of $P(s)$, we have confirmed the occurrence of
chaos in the anharmonic Jahn-Teller oscillation and the energy scale
of the chaotic region.
It has been emphasized that the chaotic behavior in the vibronic state
of dynamical Jahn-Teller system originates from the anharmonic
Jahn-Teller oscillation.
We have proposed to observe the peak structure in the specific heat
of cage-structure materials as a signal of the chaotic nature.
It is a novel possibility to detect chaos in nonlinear physics
by the standard experiment in condensed matter physics.

This work has been supported by
JSPS KAKENHI Grant Numbers 25400405 and 24540379.
The computation has been done using the facilities of
the Supercomputer Center, the Institute for Solid State Physics,
the University of Tokyo.



\begin{thebibliography}{99}
\bibitem{unit}
In this paper, we use such units as
$\hbar$=$k_{\rm B}$=$1$.
\end{thebibliography}
\end{document}